\documentclass[conference]{IEEEtran}
\IEEEoverridecommandlockouts
\usepackage[hidelinks]{hyperref}
\usepackage{cite}
\usepackage{amsmath,amssymb,amsfonts}
\usepackage{algorithmic}
\usepackage{graphicx}
\usepackage{textcomp}
\usepackage{xcolor}
\usepackage{mathptmx}
\usepackage{microtype}
\usepackage{paralist}
\usepackage{booktabs}
\usepackage{balance}
\usepackage{ragged2e}
\usepackage{siunitx}
\newcolumntype{d}{S[table-format=3.2]}

\usepackage{algorithm2e}

\DontPrintSemicolon
\SetKwComment{tcp}{$\triangleright$ }{}
\SetVlineSkip{0cm}
\SetKwInOut{Param}{Param}
\SetKwInOut{Input}{Input}

\SetKwFor{Parfor}{for}{do in parallel}{endfor}

\renewcommand\_{\textunderscore\allowbreak}

\makeatletter
\newcommand{\removelatexerror}{\let\@latex@error\@gobble}
\makeatother

\newcommand{\Neig}{\operatorname{Neig}}

\newcommand{\doubleblind}[1]{#1}

\begin{document}

\title{Concurrent Graph Queries on the Lucata Pathfinder }

\author{\IEEEauthorblockN{\doubleblind{Emory Smith}}
\IEEEauthorblockA{\doubleblind{\textit{Lucata}} \\
 \doubleblind{esmith@lucata.com}}
\and
\IEEEauthorblockN{\doubleblind{Shannon Kuntz}}
\IEEEauthorblockA{\doubleblind{\textit{Lucata}} \\
\doubleblind{skuntz@lucata.com}}
\and
\IEEEauthorblockN{\doubleblind{Jason Riedy}}
\IEEEauthorblockA{\doubleblind{\textit{Lucata}} \\
\doubleblind{jason@acm.org}}
\and
\IEEEauthorblockN{\doubleblind{Martin Deneroff}}
\IEEEauthorblockA{\doubleblind{\textit{Lucata}} \\
\doubleblind{mdeneroff@lucata.com}}
}

\maketitle

\begin{abstract}
  High-performance analysis of unstructured data like graphs now is
  critical for applications ranging from business intelligence to genome
  analysis. Towards this, data centers hold large graphs in memory to
  serve multiple concurrent queries from different users. Even a single
  analysis often explores multiple options. Current computing
  architectures often are not the most time- or energy-efficient
  solutions. The novel Lucata Pathfinder architecture tackles this
  problem, combining migratory threads for low-latency reading with
  memory-side processing for high-performance accumulation. One hundred
  to 750 concurrent breadth-first searches (BFS) all achieve
  end-to-end speed-ups of 81\% to 97\% over one-at-a-time queries on a graph with
  522M edges. Comparing to RedisGraph running on a large Intel-based
  server, the Pathfinder achieves a 19$\times$ speed-up running 128 BFS queries
  concurrently. The Pathfinder also efficiently supports a mix of
  concurrent analyses, demonstrated with connected components and BFS.
\end{abstract}


\section{Introduction}

Graph databases and analysis provide the basis for
applications dealing with unstructured data coming from
business intelligence, medicine, security, and
more\cite{sakr2021future,sahu-2019-ubiquit-large}.
These require rapid non-obvious relationship
analysis\cite{Jonas_2006} across large-scale data sets.
Often graph databases are kept resident in memory; their irregular
access patterns defeat most optimizations for slower storage.

Providing enough memory, possibly across multiple systems,
incurs a substantial monetary cost.  Ideally these graph
databases would support multiple different analysis queries
running concurrently.  These can come from one source, an
analyst or program trying multiple options, or from a large
number of sources as in a web-accessible graph database.

The irregular accesses in many graph analysis algorithms also
cause problems for current computer architectures.  These are
designed primarily for \emph{dense} operations either in
relational databases or linear algebra and machine learning.
Modern processors, including general-purpose graphics
processing units (GPGPUs), optimize memory access through
aggregation in cache lines or ``warps'' and vectors.
The optimizations are detrimental when the algorithm only
needs one or two pieces of data from the collection of eight
to hundreds\cite{6567199,10.1145/3418082}.

The mis-alignment between application needs and current
architectures has lead to many new ideas and architectural
innovations. One considered in this paper is the Lucata (n\'ee
Emu Technology) architecture based on migratory threads,
narrow-channel memory, and in-memory
processing\cite{emu-2016}.
The 32-node Lucata Pathfinder used in for evaluation in
 Section~\ref{sec:evaluation} is
 ranked at 219 in the June 2022 Graph500 BFS
 results\footnote{\url{https://graph500.org/?page_id=1073}}
 and at 52 in the June 2022 Green Graph500 Big Data
 results\footnote{\url{https://graph500.org/?page_id=1077}}.
 These results are from an FPGA-based implementation with
 processor cores running at only 225\,MHz, demonstrating
 that the other platforms are bottlenecked by their memory
 systems.

This paper provides the first study of running multiple graph
analyses \emph{concurrently} on the Lucata Pathfinder, the
latest version of the Lucata architecture.  We compare
performance of multiple concurrent breadth-first searches
(BFS) from
unique source vertices with the same searches run in sequence.
These find performance improvements of 81\% to 97\% without
any explicit scheduling or allocation of resources (Section~\ref{sec:perf-conc-v}).

Running
a mix of BFS and an architecture-specific connected components
algorithm provides improvements of
70\% on an eight-node configuration and 38\%--47\% on the full
Pathfinder (Section~\ref{sec:comb-bfs-conn}).
The connected components algorithm leverages a unique feature
of the Lucata architecture to achieve high performance:
in-memory computation of integer minimum operations.

We additionally provide an initial comparison against RedisGraph, a
commercial graph database that supports concurrent queries.
Including an approximation for the client-server overhead, the Pathfinder
achieves a 19$\times$ speed-up compared to RedisGraph running on
high-end Intel Xeon servers (Section~\ref{sec:redisgraph}).

The next section provides a brief overview of the Lucata
architecture. Section~\ref{sec:graph-algorithms} references
the BFS algorithm used and explains the adaptation of the
Shiloach-Vishkin connected components algorithm\cite{sv-1982}
to the architecture.
Section~\ref{sec:evaluation} provides implementation details
and experimental results.
A brief overview of related work follows in
Section~\ref{sec:related-work}.
Finally, Section~\ref{sec:conclusion} summarizes our
observations and provides future directions for further study,

 \section{Lucata Architecture}
\label{sec:lucata-architecture}

\begin{figure}
  \centering
  \includegraphics[width=\columnwidth]{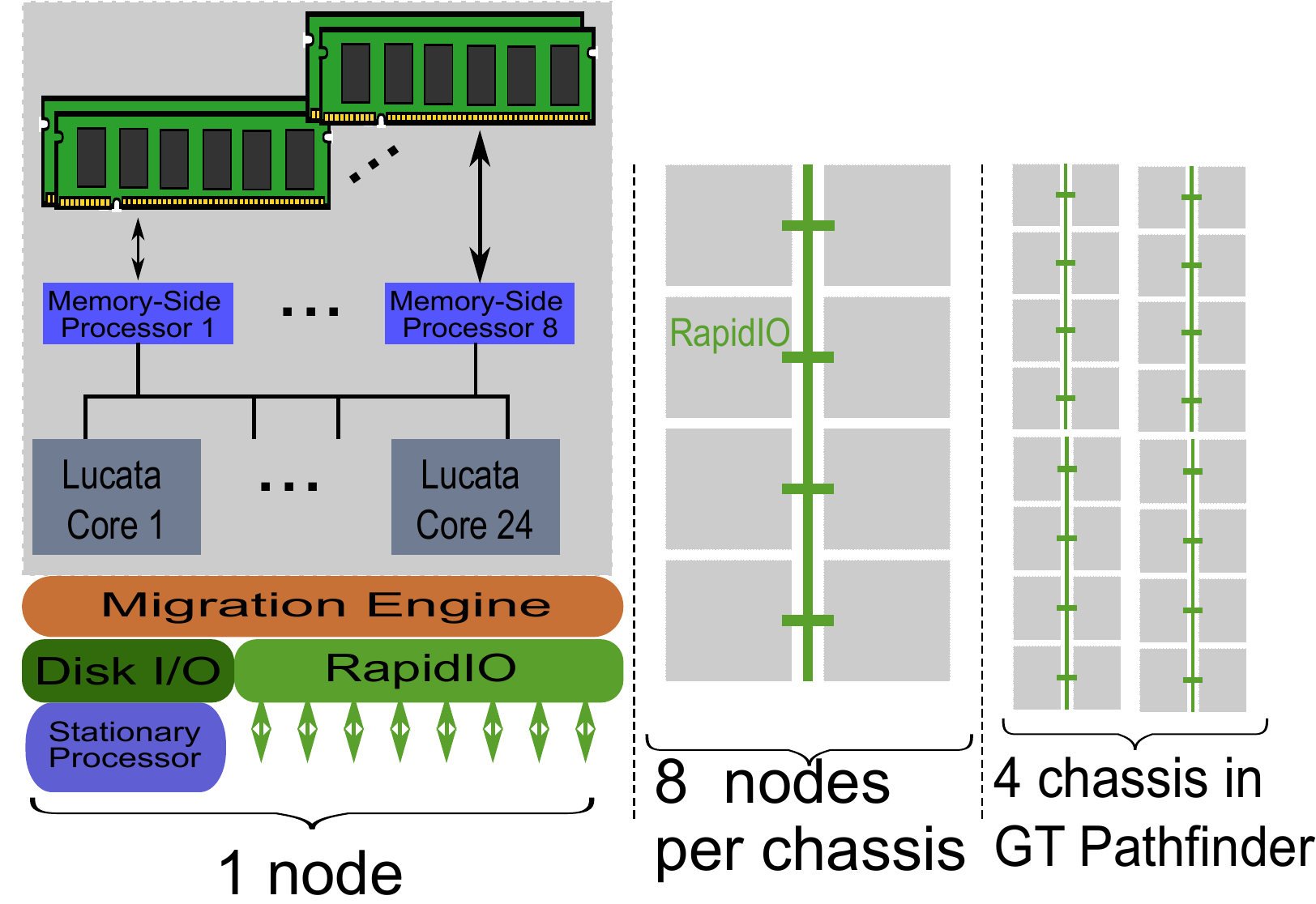}
  \caption{Lucata architecture as in the Pathfinder installed
    at GT CRNCH}
  \label{fig:lucata-arch}
\end{figure}

The Lucata architecture improves random-access memory bandwidth scalability by migrating threads to data and emphasizing fine-grained memory access.
A single Lucata Pathfinder \emph{node} consists of the following processing elements, as illustrated in Figure~\ref{fig:lucata-arch}:
\begin{itemize}
\item a \emph{stationary} processor runs the OS (Linux)
  and manages storage and network devices,
\item eight sets of banked memory coupled with an in-memory
  processor,
\item a \emph{hardware} thread migration engine, and
\item twenty-four highly multi-threaded, cache-less Lucata cores.
\end{itemize}
A Pathfinder chassis includes eight nodes.
 Each node contains 64\,GiB of narrow-channel memory; a chassis
contains 512\,GiB of memory.
These elements are combined with a RapidIO fabric
spanning the entire system.  The nodes also include SSDs for
external storage, but they are not used for these experiments
beyond storing the executable and graph data set.  Both are
loaded before any timings.

For programmers, the programs are compiled for and run on the
Lucata cores. The memory is globally addressable, forming a
partitioned global address space (PGAS).
Parallelization is expressed through
Cilk\cite{leiserson1997programming,opencilk2018} fork-join
tasks with Lucata extensions. The extensions provide a
mechanism to launch a thread and create its initial stack on
the same node as a specific address.

A launched thread only performs \emph{local reads}. Any remote
read triggers a migration which transfers the thread's context
to a processor local to the memory channel containing the
data.
This eliminates high-latency remote reads when many reads are
clustered in one memory.
The Pathfinder minimizes thread migration overhead by limiting
the size of a thread context, implementing the transfer
efficiently in hardware, and integrating migration throughout
the architecture.
Memory writes, however, do not trigger migrations and are
handled by the memory-side processors described below.

Hardware supports multiple \emph{views} of memory via fields
in the addresses beyond the 48 bits used for global physical
addresses. View zero provides a fixed address to the same
physical address on each node. These provide access to
``constants'' like the count of a graph's vertices without
migration.  View one is the global address.
View two, however, stripes 64-bit elements across nodes.  For an
address $p$ on node $n$, $p+8$ is on node $n+1$.   Arrays
based on view-one addresses are balanced across the system.


The highly multi-threaded Lucata cores read only local memory
and do not have caches, avoiding cache coherency traffic.
Each core supports 64 hardware thread contexts with
round-robin issue of one thread per cycle.  The configuration
in Figure~\ref{fig:lucata-arch} supports 1\,536 active thread
contexts per node.
Additionally, \emph{memory-side processors} (MSPs) provide a
set of remote write operations like addition that can be used
to modify small amounts of integer or floating-point data
without triggering unnecessary thread migrations.

A node's memory size is relatively large but with multiple,
Narrow-Channel DRAM (NCDRAM) memory channels. Each DIMM
has a page size of 512B and a row size of 1024. The smaller
bus means that each channel of NCDRAM has only 2GB/s of
bandwidth, but the system makes up for this by having many
more independent channels. Because of this, it can sustain
more simultaneous fine-grained accesses than a traditional
system with fewer channels and the same peak memory bandwidth.

\section{Graph Algorithms}
\label{sec:graph-algorithms}

The experiments in Section~\ref{sec:evaluation} consider two
algorithms: breadth-first search (BFS) and connected
components.
The tuned algorithm for breadth-first search
leverages thread migrations but balances migrations with
memory writes that can be remote without migrating.  Details
do not fit in this paper but are detailed fully
in~\cite{10.1145/3418077,hein2018near} .
The connected components algorithm uses a feature unique to
the Lucata architecture: computing the minimum of two integers
in-memory through the memory-side processors (MSP).

\begin{figure}
  \removelatexerror
  \centering
  \begin{algorithm}[H]
    \begin{small}
      $C[v] \leftarrow v \ \ \forall v\in V$\; changed
      $\leftarrow$ false\; \For{iter $= 0$; iter $<$
        max\_iter; $++$iter}{
        $pC[v] \leftarrow C[v] \ \ \forall v \in V$\;
        changed $\leftarrow$ false on all nodes\;
        \Parfor{$v \in V$}{
          \Parfor{$j \in \Neig(v)$}{
\lnl{remote-min}            remote\_min($\&C[j], C[v]$)\; } }
        \Parfor{$v \in V$}{ \If{$pC[v] \not= C[v]$}{changed
            $\leftarrow$ true} }
        \lnl{reduce} Reduce \emph{changed }via logical \textbf{or}\;
        \If{changed $=$ false}{break}

        \Parfor{$v \in V$}{ \While{$C[v] \not=
            C[C[v]]$}{$C[v] \leftarrow C[C[v]]$} } }
    \end{small}
  \end{algorithm}
    \caption{Connected components using the
      \textbf{remote\_min} operation via memory-side
      processors.}
  \label{lst:cc-remote}
\end{figure}
In Algorithm~\ref{lst:cc-remote}, $G = (V, E)$ is the undirected graph, where
$V$ and $E$ denote the graph's set of vertices and edges, respectively.
We assume all graphs are undirected but are represented in a
directed format.  Hence the representation contains both $(i,
j)$ and $(j, i)$ whenever either is in $E$.
The integer array $C$ holds the current component label of
each vertex.  The array $pC$ stores the labels before
performing another round of ``hooking'' or connecting tentative
components.  The operator $\Neig(v)$ denotes the set of
neighbor vertices of $v$, so the end point $j$ of all edges
$(v, j) \in E$.  We have experimented with many of the more
recent sampling-based algorithms\cite{ConnectIt,8425156} but
have yet to match the simpler algorithm's performance.
Investigating remote operations in label-propagation
algorithms\cite{9555986} is future work.

The algorithm in Figure~\ref{lst:cc-remote} essentially
describes the
Shiloach-Vishkin algorithm\cite{sv-1982} but with one Lucata-specific twist.
We use the \textit{remote\_min} operation provided by the MSP
to ``push'' the minimum labels on line~\ref{remote-min}. These
remote operations are performed while the MSP accesses the
physical memory, encapsulating the operation in a
read-modify-write cycle. Moreover the MSPs act independently
from the processing cores. The remote operations do not
interfere with local thread scheduling or execution, although
all compete for memory access. Having eight MSPs per node
eliminates most contention; we continue evaluating the
appropriate trade-offs.

While remote\_min eliminates thread migrations in the first
phase, the \emph{compress} phase that flattens the min-based
trees does migrate.  The number of migrations, however, is
bound by the depth of each tree, and that is reduced to one by
each compress phase.

Additionally, the \emph{changed} variable is kept in view-zero
storage.  Each node then may have a different value of the
changed flag.  We reduce these to a single value on the
primary node (line~\ref{reduce}) through a simple loop that
migrates a across the nodes by casting the pointer back to a
global, view-one address.

\section{Evaluation}
\label{sec:evaluation}

The Pathfinder  installed at
Georgia Tech's Center for Research into Novel Computing
Hierarchies (CRNCH)\cite{rg-icrc-2019}, includes four chassis
and a total of 2\,TiB of narrow-channel memory.  The
instruction set and other engineering parameters in the
Lucata cores is changing, so they are implemented in FPGAs.
The Pathfinder for these experiments use 24 cores per node
running at 225\,MHz.  The CRNCH Pathfinder was the first
system delivered, and the first system outside Lucata running at the four-chassis
scale.  A few hardware issues that affect performance were
identified and are described in Section~\ref{sec:perf-conc-v}.

\subsection{Implementation and Dataset Details}
\label{sec:impl-details}

We represent graphs in a loose sparse row format.
The vertex records are stored in a dense array, and each
record points to an edge block.
We use a directed representation for the undirected graph $G$;
we store both $(i, j)$ and $(j, i)$ for every edge.
The edge block is an array of neighbor vertices, storing each
$j \in \Neig(i)$ for vertex $i$.
The vertex array is striped across the system, and the edge
block is stored on the same node as the vertex's entry.  So
vertex 0 and its neighbor array is on node 0, vertex 1 and its
neighbors on node 1, and so on.  All integers are 64 bits wide.

The graph used for all experiments is generated by the
Graph500\footnote{\url{https://graph500.org}} /
R-MAT\cite{rmat-2004} generator using scale 25 and edge factor
16.  After ensuring the represented graph is undirected and removing duplicate edges, the resulting graph has
33\,554\,432 vertices and  522\,475\,613 edges.
This roughly 4\,GiB graph is relatively small for a 2\,TiB system but
permitted running experiments more rapidly.

The source vertices for BFS tests are reproducibly
pseudo-randomly generated. Running experiments multiple times
produced little enough time variation that we do not report
those results here but pick the minimum time. The variations
will depend heavily on how hardware schedules threads, and
that is an area under heavy development.

\subsection{Performance of Concurrent v. Sequential BFS}
\label{sec:perf-conc-v}

\begin{figure}[t]
  \centering
  \includegraphics[width=\columnwidth]{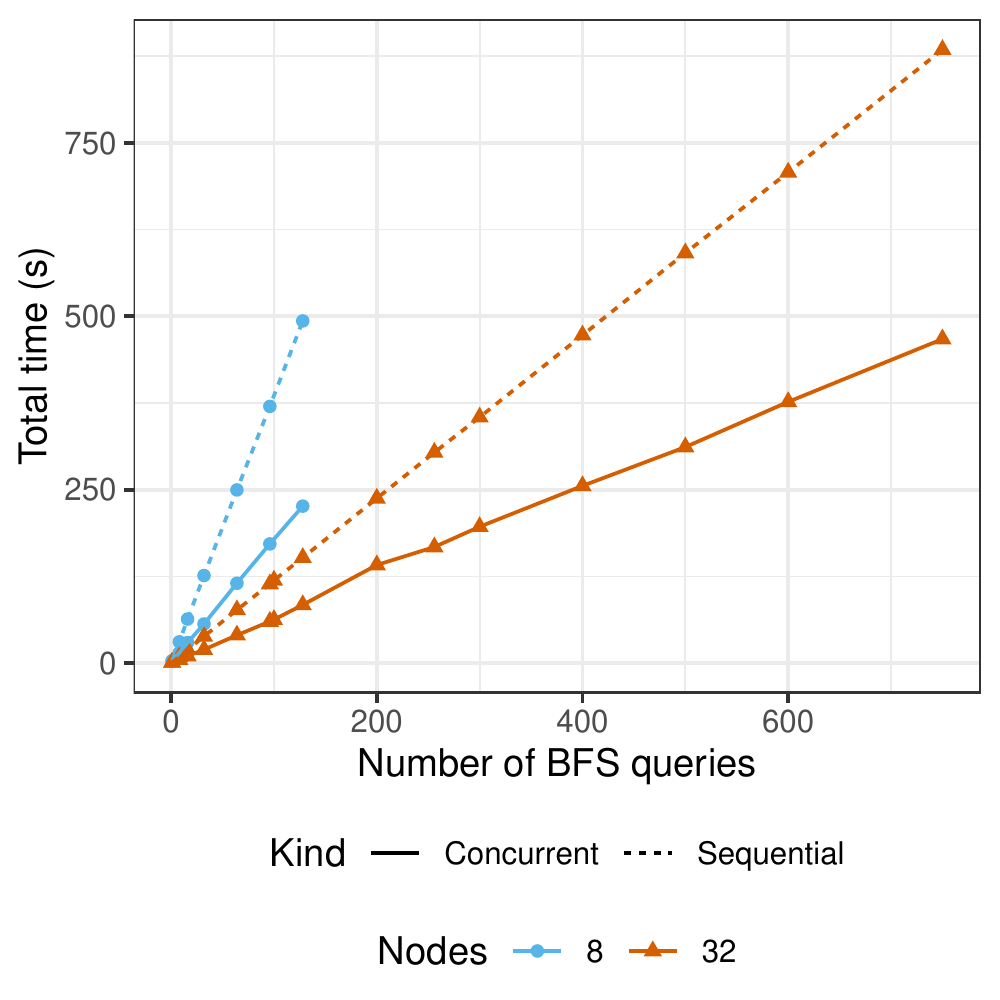}
\caption{Times (ms) for concurrent and sequential BFS
    queries}
  \label{fig:conc-bfs}
\end{figure}

Figure~\ref{fig:conc-bfs} shows the total time taken for
running unique BFS queries both sequentially and concurrently
using both eight and all 32 nodes of the CRNCH Pathfinder
system. On 32 nodes, running 750 concurrent breadth-first
searches from unique starting nodes requires 467\,s
(7.78\,min) while running the same searches one after the
other requires 884\,s (14.7\,min). On eight nodes, running 128
queries concurrently requires 226\,s (3.77\,min) while running
sequentially requires 493\,s (8.21\,min). Running 256
concurrent queries on eight nodes exhausted the memory used
for thread contexts. Further experiments will find the
boundary between the two and provide advice to users on
determining an appropriate split for their needs.

Note that the scaling from 8 to 32 nodes is \emph{not} linear.
Running the same 128 queries on both sees only a 2.69$\times$
speed-up concurrently and 3.24$\times$ speed-up sequentially.
There are multiple reasons being investigated. One is that the
128 queries may not provide sufficient parallelism for the
32-node configuration.

Another is that hardware issues with
RAM and network connections on two chassis of the four
requires reducing memory and network speed for stability. This
does not affect the chassis used for the eight-node runs. The
particular configuration also causes a two-chassis
configuration to run more slowly. Those results are not
included, but sample runs appear to require approximately
twice the time as with all four chassis.

\begin{figure}[b]
  \centering
  \includegraphics[width=\columnwidth]{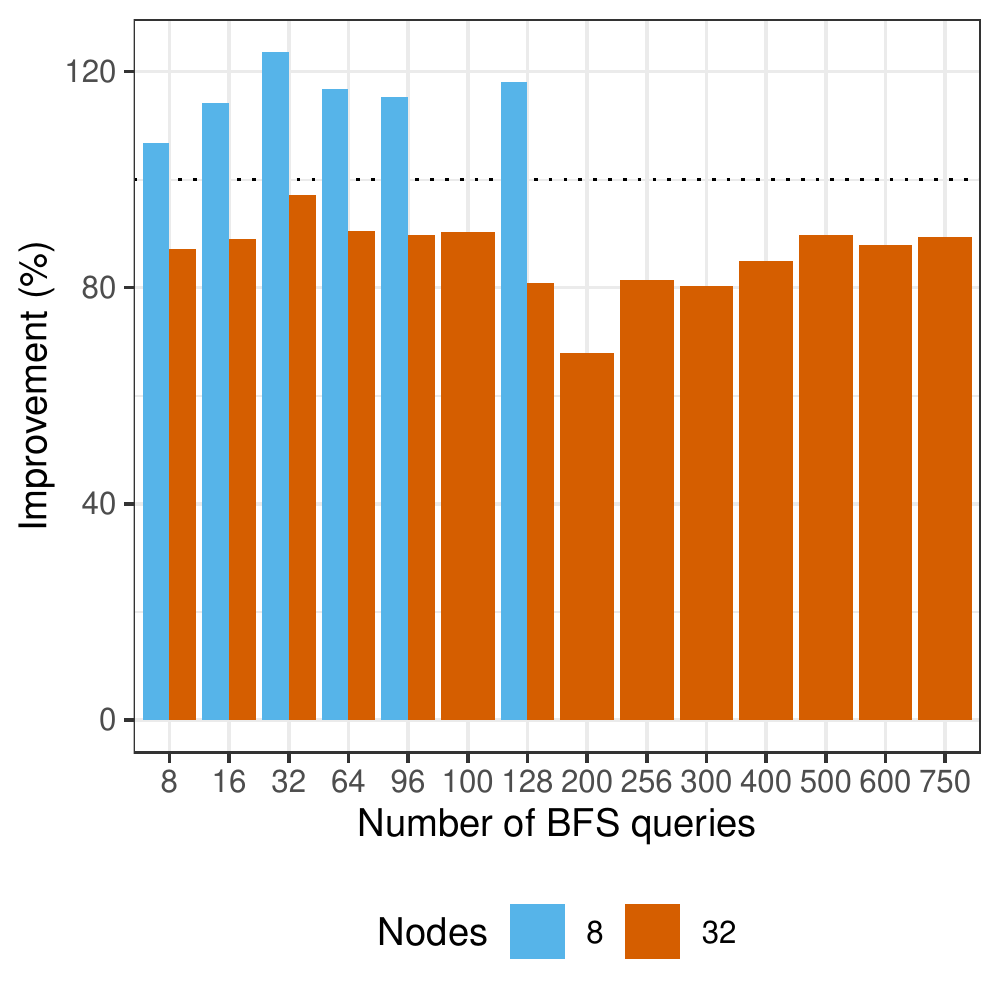}
  \caption{Improvement (\%) of concurrent queries over
    sequential queries}
  \label{fig:conc-bfs-improvement}
\end{figure}

The times increases linearly with the number of BFS
queries in all cases, but running concurrently provides a
significant improvement as shown in
Figure~\ref{fig:conc-bfs-improvement}.  The single chassis,
eight-node case consistently provides a greater than 2$\times$
speed-up of running queries concurrently rather than
sequentially.  Because of the hardware issues mentioned
previously, the four-chassis configuration does not quite
achieve the same speed-up but still runs concurrent queries
81\% to 97\% faster than sequential queries.

\begin{table}
  \centering
  \caption{Quantiles for the average time (s) per concurrent BFS}
  \label{tab:perf-quantiles}
\begin{tabular}[t]{rddddd}
\multicolumn{1}{c}{Nodes} & \multicolumn{1}{c}{0\%} & \multicolumn{1}{c}{25\%} & \multicolumn{1}{c}{50\%} & \multicolumn{1}{c}{75\%} & \multicolumn{1}{c}{100\%}\\
\midrule
8 & 1.77 & 1.80 & 2.85 & 3.87 & 3.97\\
32 & 0.61 & 0.63 & 0.94 & 1.19 & 1.22\\
\end{tabular}
 \end{table}

The improvement for concurrent queries over sequential ones
remains relatively stable across the number of queries.
Table~\ref{tab:perf-quantiles} provides the quantiles for the
average time per BFS when run concurrently.  For
eight nodes the min-max spread is 2.2\,s, while for 32 nodes the
spread is 0.61\,s.  The eight-node results contain only 12
samples of the number of concurrent queries.  The 32-node
results have 28 samples, so that 50\% of the times
lie outside the interval $[$0.63\,s, 1.19\,s$]$ is perhaps
more interesting than the total spread, nearly the same
interval.  This is worth further investigation.
\phantom{This is somewhat fishing for the reviewer explanation
  that the variation will be with fewer BFSs.}

\subsection{Combining BFS and Connected Components}
\label{sec:comb-bfs-conn}

\begin{table}
  \centering
  \caption{Times for a concurrent mix of BFS and CC, also \%
    improvement of concurrent over sequential analysis}
  \label{tab:conc-bfs-cc}
\begin{tabular}[t]{rrrddd}
\multicolumn{1}{c}{Nodes} & \multicolumn{1}{c}{\# BFS} & \multicolumn{1}{c}{\# CC} & \multicolumn{1}{c}{Conc. time (s)} & \multicolumn{1}{c}{Seq. time (s)} & \multicolumn{1}{c}{\% Impr.}\\
\midrule
8 & 136 & 34 & 649.94 & 1105.36 & 70.07\\
8 & 153 & 17 & 470.01 & 802.49 & 70.74\\
32 & 560 & 140 & 1690.85 & 2334.73 & 38.08\\
32 & 630 & 70 & 1029.25 & 1511.47 & 46.85\\
\end{tabular}
\end{table}

We expect that many to most uses will run a mix of queries concurrently.
Table~\ref{tab:conc-bfs-cc} provides the concurrent and
sequential times for running 80\%-20\% and 90\%-10\% mixes of
BFS and connected components.  These particular mixes were
chosen somewhat to balance the execution times of single
evaluations of each.  The combination of the read-heavy BFS
and the remote\_min heavy connected component algorithms also
led to some system instability at higher details.  This
currently is under heavy investigation but likely is because
of the relative priorities of read and write (like
remote\_min) operations at the memory-side processors.

As with pure-BFS, concurrent execution of the mixed queries
provides a significant improvement. Concurrently running
queries on the eight-node, single chassis reliably performs
70\% better than sequentially running all the breadth-first
searches followed by all the connected components evaluations.
There are no caches, so the equivalent connected component
queries do not benefit from any data pre-loading.

The four chassis results do not achieve as much improvement,
performing concurrent queries from 38\% to 47\% faster than
running them sequentially.  Our current hypothesis is that the
hardware issues mentioned in Section~\ref{sec:perf-conc-v} are
exacerbated by the increased global memory traffic from the
connected components algorithm's remote writes.  The recently implemented
hardware performance counters should provide more insight.

\subsection{Comparison with RedisGraph}
\label{sec:redisgraph}

Another available platform, RedisGraph\cite{redisgraph-2019} running on Intel Xeon
systems, also provides concurrent queries.  RedisGraph is
based on modified implementations of
GraphBLAS\cite{DBLP:conf/ipps/BrockBMMM21,10.1145/3322125}
and LAGraph\cite{8778338} .
We compare
the same BFS queries using RedisGraph with the Pathfinder,
although we modify the Pathfinder times to approximate
overhead in the RedisGraph parsing and client-server interface
as specified below.

The hardware platform used for Redis Enterprise is an
instance of a x1e.32xlarge Amazon Web Service (AWS) server
running Red Hat Enterprise Linux 8.6.
The server uses Intel Xeon E7-8880v3 2.3\,GHz processors to
provide 128 vCPUs, or 64 cores with 2 threads per core, and
4\,TiB of memory. RedisGraph's work pool is set to 128
threads, the maximum number of hardware contexts available to
these vCPUs. The experiments all use less than 2\,TiB of
memory, the same as available on the Pathfinder.
Redis Enterprise is version 2.6.6, and the RedisGraph module
is version 2.8.

\begin{figure}
  \centering
  \texttt{GRAPH.QUERY scale25 "MATCH(n) WHERE id(n) = 2436375
    CALL algo.BFS(n, 0, NULL) YIELD nodes RETURN size(nodes)"}

  \caption{RedisGraph query used where \texttt{id(n)}
    indicates the source vertex}
  \label{fig:rg-query}
\end{figure}

To run concurrent queries with RedisGraph, we run multiple
instances of \texttt{redis\_cli}.
This was found to be much faster than using a C interface
library not provided by Redis.
Figure~\ref{fig:rg-query}
shows a specific query; the query uses RedisGraph's built-in BFS
routine.

Using \texttt{redis\_cli} imposes quite some overhead in
parsing and communication with the server, although much of
that overhead itself overlaps across the concurrent
\texttt{redis\_cli} invocations. We attempt to adjust the
Pathfinder results that do not include equivalent overhead to provide a more
reasonable comparison.  Our assumption is that the single
\texttt{redis\_cli} instance provides a reasonable
approximation to the overhead, and we add that to all the
Pathfinder results when computing time ratios.

\begin{table}
  \caption{Time (s) and adjusted speed-ups for concurrent BFS
    queries in RedisGraph Enterprise (RG) and on the Lucata
    Pathfinder}
  \label{tbl:redisgraph}
\begin{tabular}{cdddddd}
 & 1 & 8 & 16 & 32 & 64 & 128\\
\midrule
RedisGraph & 5 & 40 & 139 & 276 & 610 & 1707\\
8 nodes & 3.47 & 14.88 & 29.69 & 56.51 & 115.21 & 226.30\\
32 nodes & 1.04 & 5.00 & 10.29 & 19.61 & 40.30 & 84.04\\
  \midrule
  \multicolumn{2}{c}{Adjusted speed-ups}\vspace*{1ex}\\
8 nodes & 0.590 & 2.01 & 4.01 & 4.49 & 5.07 & 7.38\\
32 nodes & 0.828 & 4. & 9.09 & 11.2 & 13.5 & 19.2\\
\end{tabular}
\end{table}

Table~\ref{tbl:redisgraph} provides both the times for
concurrent BFS queries and adjusted speed-ups of 8- and
32-node Pathfinder configurations.  The RedisGraph times increase
approximately linearly until running more than 32 concurrent
queries.  This AWS server does not support enough hardware thread
contexts to provide meaningful data beyond 128 queries, so we
cannot check further scalability.  Additionally, some of the
threads will be preempted for other tasks like keeping the
client-server connections alive.

In the 16- to 32-query range, the 32-node Pathfinder provides
around a 10$\times$ adjusted speed-up over RedisGraph running
on a higher-end Intel Xeon. The single-chassis, 8-node
configuration provides an adjusted speed-up of around
5$\times$. When running 128 concurrent queries, which almost
certainly goes beyond the Xeon's 128 hardware thread contexts,
the Pathfinder's speed-up jumps to 19$\times$.
Section~\ref{sec:perf-conc-v} shows that the Pathfinder's
total time increases linearly up to 750 concurrent queries on the
full system, far beyond the AWS server. A single chassis with
FPGA-implemented processors running at 225\,MHz still
outperforms the AWS server, providing some evidence for the
the Pathfinder memory system's benefits and its ability to
keep the memory busy.

\section{Related Work}
\label{sec:related-work}

\balance

Beyond RedisGraph, there is a decided uptick in work on concurrent graph database
queries and analysis. The Congra\cite{congra-2017} and
CongraPlus\cite{8642367} systems schedule to optimize usage of
typical shared-memory systems assigning CPU subsets to each.
They find limited thread scalability on the individual graph
analyses because of the limited memory access, so restricting
parallelism within each concurrent query provides higher total throughput.

For distributed systems, Wukong\cite{wukong-2021} and
Wukong+G\cite{wukong-g-2022} implement
concurrent queries on an RDF database using remote direct memory access (RDMA),
the latter across GPUs and
GPUDirect\footnote{\url{http://docs.nvidia.com/cuda/gpudirect-rdma/}}.
RDMA provides efficient remote reads but at large
granularities.  These platforms partition the graph across
memory spaces through an algorithm designed for skewed
degrees\cite{powerlyra-2019}.  Then they
fetch chunks of the graph for
local exploration.

GraphX\cite{10.1145/2484425.2484427} runs on the Spark
data-parallel framework and provides fault-tolerant
implementations of multiple ``vertex-centric''
abstractions.  Apache Giraph\cite{ching2013scaling} provides
similar capabilities atop Hadoop.  Both can support concurrent
queries through their underlying frameworks, but the cost of
fault-tolerance and the implementation technologies is
substantial.

Multiple data structures have been designed to support
concurrent access with \emph{streaming graph modifications}.
STINGER\cite{STINGER-2012} is an early example that
alternates between concurrent queries and graph updates.
Later work extended STINGER to support some graph analysis
kernels concurrently to the graph
updates\cite{hpec19-yin,yin2018new}.
Recent work on functional data structures in
Aspen\cite{Aspen-2019} support efficient snapshot views of a
changing graph for multiple analysis kernels.

\section{Conclusions and Future Work}
\label{sec:conclusion}

Running 750 multiple, concurrent breadth-first searches on the
CRNCH Pathfinder provides a speed-up of 1.9$\times$ over
running the same queries one after the other. The Pathfinder
does not include caches, and its cores run at a relatively
slow 225\,MHz. The speed-up almost certainly comes from the
large number of memory controllers / memory-side processors as
well as the ability of its highly multi-threaded cores to keep
the memory channels busy. This is a substantial improvement
over an earlier generation, the Emu Chick, which often was
limited by its inability to generate enough memory
operations\cite{10.1145/3418077}. The average time per
concurrent search shows moderate variation. The time for 128
searches does not scale perfectly from eight to 32-nodes,
likely because of hardware issues in the first external
Pathfinder.

When running a mix of breadth-first searches and connected
components queries, the full Pathfinder provides 38\% to 47\%
better end-to-end times for concurrent queries than the same
queries run in sequence. A single-chassis configuration
provides a 70\% improvement for fewer total queries. The two
algorithms have markedly different memory characteristics.
Each search level of the BFS implementation performs memory
work proportional to the size of that level, and the size
varies widely in this Graph500 dataset\cite{6468458}. The
connected components algorithm performs a remote\_min
write-style operation in the controller nearest every edge
endpoint on each of its fewer iterations. Mixing the two
stresses the interconnect and the MSPs' read/write priority
balance.

A rough comparison with RedisGraph Enterprise on a large AWS
server shows the Pathfinder performing significantly better up
to 64 concurrent breadth-first searches.  At 128 searches, the
Xeon-based server performs far worse than with fewer searches,
while the Pathfinder keeps scaling predictably.  This
comparison does not fully account for the differences in
overhead, however.

This study's experiments and methods can lead to Pathfinder
improvements, including
\begin{itemize}
\item appropriate sizing of the in-memory thread context
  reservations relative to the application data,
\item evidence for advise or heuristics for balancing
  read/write priorities in the MSPs,
\item tools for evaluating different numbers of cores and
  clock speeds, and
\item additional system diagnostics for identifying
  under-performing memory or network links.
\end{itemize}

Recently, the Pathfinder has gained hardware performance
counters.  Future work will use these to drill down into the
timings and provide evidence for some of our hypotheses.
Additionally, the counters should help diagnose the variance
in average time per BFS seen in
Table~\ref{tab:perf-quantiles}.

Because the implementation primarily uses Cilk with only a few
Lucata extensions, we also intend to compare with a direct
translation to OpenCilk\cite{opencilk2018} on traditional
architectures. This will provide a more equivalent comparison
between architectures than our RedisGraph experiments. This
requires some thought on mapping remote memory operations like
remote\_min to traditional architectures. A straight-forward
method would use standard atomic operations that well could
give the current Pathfinder architecture an ``unfair''
advantage over traditional NUMA machines, not to mention the
upcoming new architecture version.

\section*{Acknowledgments}
\addcontentsline{toc}{section}{Acknowledgments}

We thank Redis for use of RedisGraph enterprise for our study.
The RedisGraph AWS experiments were run by Internet
Infrastructure Services Corporation for the authors.
This material is based on work supported by the National
Science Foundation through grant \#2105977.
Plots use a colorblind-aware palette from Masataka Okebe and
Kei Ito\footnote{\url{https://jfly.uni-koeln.de/color/}}.

\clearpage
\bibliographystyle{IEEEtran}
\bibliography{IEEEabrv,main.bib}

\clearpage\nobalance\RaggedRight

\end{document}